\documentclass[A4,aps,twocolumn]{revtex4}
\usepackage [dvips]{graphicx}
\usepackage{multirow,amsmath,ulem}
\usepackage{color}

\begin{document}

\title{The neutron star inner crust and symmetry energy}

\author{Fabrizio Grill}
\affiliation{Centro de F\'{i}sica Computacional, Department of Physics,
University of Coimbra, P-3004-516 Coimbra, Portugal}
\author{Sidney S. Avancini}
\affiliation{Depto de F\'{\i}sica - CFM - Universidade Federal de Santa
Catarina  Florian\'opolis - SC - CP. 476 - CEP 88.040 - 900 - Brazil}
\author{Constan\c{c}a Provid\^encia}
\affiliation{Centro de F\'{i}sica Computacional, Department of Physics,
University of Coimbra, P-3004-516 Coimbra, Portugal}

\begin{abstract}
The cell structure of clusters in the inner crust of a cold
$\beta$-equilibrium neutron star is studied within a Thomas Fermi
approach and compared with other approaches which include shell
effects. Relativistic nuclear models are considered. We conclude
that the symmetry energy slope $L$ may have quite dramatic effects
on the cell structure if it is very large or  small. Rod-like and
slab-like pasta clusters have been obtained in all models except
one with a large slope $L$.

\end{abstract}
\maketitle

\section{Introduction}
The inner crust of a neutron star lies between the neutron drip
density (${\rho_d\approx 3\times10^{-4}\ fm^{-3}}$), defined as
the density where the neutrons start to drip out from the nuclei
of the crust, and the crust-core transition density (${\rho\approx
8\times10^{-2}\ fm^{-3}}$). In this region matter is most probably
formed by a lattice of heavy and neutron rich nuclei immersed in a
sea of superfluid neutrons and ultrarelativistic electrons.
Complex structures (e.g. rods and slabs) are expected to be formed
in the bottom part of the inner crust, the so-called pasta phase
region, where the transition to the homogeneous core matter
occurs.

The first microscopic calculation of the inner crust structure was
performed by Negele and Vautherin in 1973~\cite{NV}. In this work,
which is still used as a benchmark in neutron star calculations,
the inner crust was studied in the Wigner-Seitz approximation,
which divides the lattice in independent spherical cells each with
the nucleus in the center surrounded by the electron and neutron
gases. Nuclear matter was described with the Hartree-Fock (HF)
approximation based on the Density Matrix Expansion~\cite{DME}.
The parameters of that model were adjusted to reproduce the
experimental binding energies of atomic nuclei and theoretical
calculations of infinite neutron matter. The inner crust matter
calculated in that work has for each cell a magic or semi-magic
proton number (i.e. $Z=40$ and $Z=50$), indicating that in these
calculations there are strong proton shell effects, as in isolated
atomic nuclei.

Later Douchin and Haensel proposed a new study where the neutron
star structure was calculated using the same Equation of State
(EOS) in the whole star, from the outer crust to the
core~\cite{Haensel}. In that work nuclear matter was treated in
the Wigner-Seitz (WS) approximation using the non-relativistic
Compressible Liquid Drop (CLD) model and the SLy4 effective
interaction~\cite{sly4}. The shell effects were neglected both for
protons and neutrons. In the transition between the inner crust
and the core five different pasta shapes were considered:
spherical nuclei (droplets), cylinders of nuclear matter (rods),
and plane slabs of nuclear matter (slabs) in a neutron gas, as
well as cylindrical holes (tubes) and spherical holes (bubbles) in
nuclear matter filled with a neutron gas. Within that model these
authors found that in the whole inner crust the shape that
minimizes the energy is the droplet and that those cells were
characterized by a weak change in the proton number: from
$Z\approx 40$ near the neutron drip to $Z\approx 50$ in the region
of the transition to the core. Moreover because of the absence of
proton shell effects in all the calculated configurations the
proton number was not magic or semi-magic.

The effect of pairing correlations on the structure of WS cells
was investigated for the first time in~\cite{baldo} within the
Hartree-Fock-Bogoliubov approach (HFB). In the most recent version
of these calculations these authors solved the HFB equations using
an equation of state (EOS) mixture of the phenomenological
functional of Fayans et al~\cite{fayans}, employed in the nuclear
cluster region, and a microscopical functional derived from
Bruckner-Hartree-Fock calculations in infinite neutron matter. In
this framework it was found that the cells have not a magic or
semi-magic number of protons and that pairing can change
significantly the structure of the cells compared to HF
calculations.

The pairing effect was also studied by G\"{o}gelein and M\"{u}ther
in 2007~\cite{Gogelein}. In this work these authors used a
self-consistent Skyrme-Hartree-Fock (SHF) approach in order to
study the cell structure in the pasta phase region (${0.01\
\mbox{fm}^{-3}<\rho<0.1\ \mbox{fm}^{-3}}$), where three different
pasta shapes were considered: droplet, rod and slab. The pairing
correlations were evaluated within the BCS approach, assuming a
density-dependent contact interaction, while, for the nuclear
interaction, the SLy4 parametrization~\cite{sly4} was used.
Contrary to what was obtained in~\cite{Haensel} in this work the
calculations showed that all the three cell shapes appear in the
inner crust before the transition to the homogeneous core.
Moreover the calculated core-crust transition density was coherent
with the value that was found in~\cite{Isaac} for the SLy4
interaction within a dynamical spinodal calculation.

Recently Grill et al. proposed a new inner crust structure
calculation based on the HFB approach~\cite{grill}. This study was
performed in the regions which are supposed to be formed by a
lattice of spherical clusters. Thus the inner crust matter was
divided in spherical cells treated in the Wigner-Seitz
approximation and the structure of these cells (i.e. neutron
number (N), proton number (Z) and cell radius ($R_{ws}$)) were
obtained from the energy minimization at beta equilibrium. For the
HFB calculations it was considered a SLy4 interaction~\cite{sly4}
in the particle-hole channel, while in the particle-particle
channel three zero-range density-dependent pairing forces of
various intensities were used. With this model it was possible to
find very reliable results in the low density regions of the inner
crust. Indeed the calculated structure is coherent with the
results of outer crust calculations in the literature
(e.g.~\cite{RocaOut}). Moreover, in those regions a very weak
dependence on the pairing interaction was found. On the other hand
in the high density regions of the inner crust the imposed
discretization of the free neutron gas generates an
underestimation of the energy in the smaller cells: it was thus
used an empirical correction~\cite{corr} which, however, was
characterized by large fluctuations, so the structure of the cells
in those regions was not guaranteed by those calculations.

In this paper we present a calculation of the inner crust
structure using a relativistic mean field density dependent
Thomas-Fermi approach (TF)~\cite{Sidney}. A complete self
consistent calculation is performed, namely, no parametrization of
the density and surface energy as in~\cite{Gogelein} has been
used. This approach, which neglects the shell effect for both
neutrons and protons, is less accurate than the HFB in the low
density regions, where the nucleus influences the cell properties.
However, in the high density regions, when the free neutron gas
becomes more important than the nucleus, the accuracy of the TF
approximation can be higher than that of the HFB approach: indeed
with this model the discretization of the free neutron gas is not
necessary. Moreover, with the HFB approach it is actually possible
to consider just the spherical symmetry and so the droplet is the
only available shape of the cells. On the other hand with the TF
approach it is possible to study also the cylindrical and the
plane symmetries, making available also the rod and slab cell
shapes: we have performed our study also in the pasta phase region
and determined the cell structure until the crust-core transition.

 The neutron inner crust is particulary
sensitive to the density dependence of the symmetry energy \cite{oyamatsu07},
and, therefore, comparing results obtained with different nuclear interactions
will show how the symmetry energy affects the cell structure.
This comparison is possible within a TF but  would be
prohibitive within a HFB calculation due to the excessive CPU time required.

\section{Formalism}
We will apply in the present study the self-consistent
Thomas-Fermi calculation presented in~\cite{Sidney,Sidney2},
within relativistic nuclear models with constant couplings and
non-linear terms~\cite{bb}, and with density dependent
couplings~\cite{tw}. In the relativistic mean-field theory protons
and neutrons interact with and through an
isoscalar-scalar field $\sigma$, an
isoscalar-vector field $\omega^{\mu}$, an
isovector-vector field $\boldsymbol{\rho}^{\mu}$  and an isovector-scalar field $\boldsymbol{\delta}$. Within the first class of models, that we will
designate by Non Linear Walecka Models (NLWM), we consider
NL3~\cite{nl3} with non linear $\sigma$ terms, NL3$_{\omega\rho}$
including also non-linear $\omega\rho$ terms which allow the
modulation of the density dependence of the symmetry
energy~\cite{hor01}, FSU~\cite{fsu} and IU-FSU~\cite{iufsu} with
non-linear $\sigma$, $\omega$ and $\omega\rho$ terms. These two
parametrizations were constrained by the collective response of
nuclei to the isoscalar monopole giant resonance (ISGMR) and the
isovector dipole giant resonance (IVGDR).
Within the second class of models with density dependent couplings
we consider DD-ME2~\cite{ddme2} and DD-ME$\delta$~\cite{roca2011}:
DD-ME2, as all the non-linear parametrizations considered, does not
include the $\delta$ meson, and was adjusted to experimental data
based on finite nuclei properties; DD-ME$\delta$ contains the
$\delta$ meson and was fitted to microscopic ab-initio
calculations in nuclear matter and finite nuclei properties.

Stellar matter will be described by a mixture of protons, neutrons
and electrons in chemical equilibrium. Electrons
are described as a relativistic fermion gas which interacts with
protons through the electromagnetic field $A^{\mu}$.

All the equations that allow the performance of the Thomas-Fermi
calculation are derived from the Lagrangian density
\begin{equation}
\mathcal{L}=\sum_{i=p,n}\mathcal{L}_{i}+\mathcal{L}_e\mathcal{\,+L}_{{\sigma }}%
\mathcal{+L}_{{\omega }}\mathcal{+L}_{{\rho
}}\mathcal{+L}_{{\delta}}\mathcal{+L}_{{\gamma }}
\mathcal{+L}_{{nl}}, \label{lagdelta}
\end{equation}
where the nucleon Lagrangian reads
\begin{equation}
\mathcal{L}_{i}=\bar{\psi}_{i}\left[ \gamma _{\mu }iD^{\mu }-M^{*}\right]
\psi _{i}  \label{lagnucl},
\end{equation}
with
\begin{eqnarray}
iD^{\mu } &=&i\partial ^{\mu }-\Gamma_{\omega}\Omega^{\mu }-\frac{\Gamma_{\rho }}{2}{\boldsymbol{\tau}}%
\cdot \boldsymbol{\rho}^{\mu } - e \frac{1+\tau_3}{2}A^{\mu}, \label{Dmu} \\
M^{*} &=&M-\Gamma_{\sigma}\sigma-\Gamma_{\delta
}{\boldsymbol{\tau}}\cdot \boldsymbol{\delta}, \label{Mstar}
\end{eqnarray}
and the electron Lagrangian is given by
\begin{equation}
\mathcal{L}_e=\bar \psi_e\left[\gamma_\mu\left(i\partial^{\mu} + e A^{\mu}\right)
-m_e\right]\psi_e.
\label{lage}
\end{equation}
The meson and electromagnetic Lagrangian densities are
\begin{eqnarray*}
\mathcal{L}_{{\sigma }} &=&\frac{1}{2}\left( \partial _{\mu }\sigma \partial %
^{\mu }\sigma -m_{\sigma}^{2}\sigma ^{2}\right)  \\
\mathcal{L}_{{\omega }} &=&\frac{1}{2} \left(-\frac{1}{2} \Omega _{\mu \nu }
\Omega ^{\mu \nu }+ m_{\omega}^{2}\omega_{\mu }\omega^{\mu } \right) \\
\mathcal{L}_{{\rho }} &=&\frac{1}{2} \left(-\frac{1}{2}
\mathbf{R}_{\mu \nu }\cdot \mathbf{R}^{\mu
\nu }+ m_{\rho }^{2}\boldsymbol{\rho}_{\mu }\cdot \boldsymbol{\rho}^{\mu } \right)\\
\mathcal{L}_{ {\delta }} &=&\frac{1}{2}(\partial _{\mu }\boldsymbol{\delta}%
\partial ^{\mu }\boldsymbol{\delta}-m_{\delta }^{2}{\boldsymbol{\delta}}^{2})\\
\mathcal{L}_{{\gamma }} &=&-\frac{1}{4}F _{\mu \nu }F^{\mu
  \nu }\\
\mathcal{L}_{{nl}} &=&-\frac{1}{3!}\kappa \sigma ^{3}-\frac{1}{4!}%
\lambda \sigma ^{4}+\frac{1}{4!}\xi g_{\omega}^{4}(\omega_{\mu}\omega^{\mu })^{2} \\
&+&\Lambda_\omega \Gamma_\omega^2 \Gamma_\rho^2 \omega_{\mu
}\omega^{\mu } \boldsymbol{\rho}_{\mu }\cdot
\boldsymbol{\rho}^{\mu }
\end{eqnarray*}
where $\Omega _{\mu \nu }=\partial _{\mu }\omega_{\nu }-\partial
_{\nu }\omega_{\mu }$, $\mathbf{R}_{\mu \nu }=\partial _{\mu
}\boldsymbol{\rho}_{\nu }-\partial _{\nu }\boldsymbol{\rho}_{\mu
}-\Gamma_{\rho }(\boldsymbol{\rho}_{\mu }\times
\boldsymbol{\rho}_{\nu })$ and $F_{\mu \nu }=\partial _{\mu
}A_{\nu }-\partial _{\nu }A_{\mu }$. The four coupling parameters
$\Gamma_\sigma$, $\Gamma_\omega$, $\Gamma_{\rho}$ and
$\Gamma_{\delta}$ of the mesons to the nucleons are density
dependent in the relativistic density dependent models considered,
namely, DD-ME2~\cite{ddme2} and DD-ME$\delta$~\cite{roca2011}. The
non-linear term $\mathcal{L}_{{nl}} $ is absent in these models.
In all the other models, NL3~\cite{nl3},
NL3$_{\omega\rho}$~\cite{hor01}, FSU~\cite{fsu} and
IU-FSU~\cite{iufsu}, the couplings are constants and at least some
of the non-linear terms of $\mathcal{L}_{nl}$ are included. In the
above Lagrangian density $\boldsymbol {\tau}$ is the isospin
operator.

The results obtained within the above relativistic mean-field
models will be compared with the corresponding results calculated
with the non-relativistic effective Skyrme interaction
SLy4~\cite{sly4} within two formalisms: a HFB
calculation~\cite{grill} and a CLD calculation~\cite{Haensel}. We
will discuss how sensitive is the structure of the non-homogeneous
inner-crust of a neutron star to the properties of the EOS and the
formalism used. In Table~\ref{TabSat} the saturation nuclear
matter properties and the $\sigma$ meson mass are shown. The
$\sigma$ meson mass has a strong influence on the nuclear surface
energy and is included in the table to help the discussion.

In Fig.~\ref{EsymFig} the symmetry energy and its slope $L$ are
plotted for all the models. The models considered have very
similar values for the symmetry energy at saturation, namely,
between 31.3 and 32.6 MeV except NL3 that has a quite high value,
37.3 MeV. However, there is a larger dispersion of the symmetry
energy slope $L$ with values between 45 and 60 MeV, together with
118 MeV for NL3. All models behave in a similar way except NL3
whose slope is much larger above $\rho> \rho_0/3$. SLy4 has the
smallest slope $L$ only at saturation density. FSU has the second
largest slope only above the density $\rho\sim 0.7 \rho_0$. The
properties of the pasta will reflect these facts, with IU-FSU and
NL3 behaving in a quite different way, while all the other models
showing similar results. The slope $L$ has a direct influence on
the surface energy tension and surface thickness of the clusters.

A smaller $L$ corresponds generally to a larger surface tension
and smaller neutron skin thickness~\cite{hor01} as can be
confirmed comparing the surface tensions of the above models. In
Fig.~\ref{sigma} the surface tension obtained from the derivatives
of the meson fields, as indicated in~\cite{Sidney3}, are plotted.
The main differences between the models are due to the properties
of the EOS at subsaturation densities: a smaller $L$ dictates a
larger surface tension in asymmetric matter: IU-FSU has the
smallest $L$ and largest surface tension; it is this effect that
explains the difference between NL3 and NL3$_{\omega\rho}$. The
dependence of the surface tension in symmetric nuclear matter on
the model properties was well discussed in~\cite{estal99}, a
smaller incompressibility will give a smaller surface tension:
this explains the small DD-ME$\delta$ value; a smaller value of
$m_\sigma$ and a larger saturation density will give rise to a
larger surface tension as in IU-FSU.

\section{Results}

In our study of the inner crust structure we have considered three
different cell shapes: the droplet, the rod and the slab. In most
part of the inner crust, that we will designate by standard inner
crust~(${\rho\leq 4.71\times10^{-2}\ \mbox{fm}^{-3}}$),
we have limited our study just to the droplet configuration, so,
as in~\cite{NV,grill}, the lattice structure is described as a set
of spherical cells, with radius $R_{ws}$, treated in the
Wigner-Seitz approach. The volume of a spherical cell with radius
$R_{ws}$ is
\begin{equation}\label{VolD}
 V^{d}(R_{ws})=\frac{4}{3}\pi R_{ws}^3 .
\end{equation}
However in the higher density regions, that we will designate by pasta
phase regions~(${\rho\geq 4.71\times10^{-2}\ \mbox{fm}^{-3}}$), we have taken into account
all the three shapes. Rod cells have cylindrical shape with the
radius $R_{ws}$ and length set for simplicity to $l=30$ fm, which is $\sim
1.5\, R_{ws}$ in region 1 (the
final results are independent of this parameter)
\begin{equation}\label{VolR}
 V^{r}(R_{ws})= l \pi R_{ws}^2 .
\end{equation}
Slab cells have the shape of a parallelepiped with width and
length set to $l=30$ fm (the final results are independent on this
parameter, too) and depth equal to $2R_{ws}$, so that
\begin{equation}\label{VolS}
 V^{s}(R_{ws})= 2l^2 R_{ws} .
\end{equation}

We consider these three different cell shapes at a given density
and calculate the cell structure through the minimization of the
energy per baryon under the condition of $\beta-$equilibrium. For
a fixed number of protons and neutrons at a given density it is
possible to obtain univocally the cell radius ($R_{ws}$)
\begin{equation}\label{FixDen}
    V^i(R_{ws})=\frac{N+Z}{\rho}\hspace{1 cm}i=d,r,s ,
\end{equation}
and by the $\beta-$equilibrium condition for each proton number it
is possible to calculate the neutron number. Finally the cell
structure is univocally defined searching the cell shape and the
proton number that minimize the energy per baryon. In this
minimization process we have treated $Z$ and $N$ as an integer
(contrary to the approach of Douchin and Haensel~\cite{Haensel}).
The main effects of imposing this condition are: smaller
Wigner-Seitz cells and larger energies are obtained.

In particular for each fixed density we have calculated the cell
shape, the proton and neutron number, the cell radius ($R_{ws}$),
which is defined from the cell volume~(\ref{VolD}), (\ref{VolR}) and
(\ref{VolS}), the energy per baryon (${E/A}$), the neutron chemical
potential ($\mu_N$) and the proton fraction (${x=Z/(Z+N)}$).

\subsection{Standard inner crust}

The properties of the standard inner crust are reported in the
Tables~\ref{TabZNRw} and \ref{TabEMuX} where the proton number,
neutron number, Wigner-Seitz radius, energy per baryon, neutron
chemical potential and total proton fraction defined for twelve
different values of the density defined in the second line of
these tables. The first ten were introduced by Negele and
Vautherin in \cite{NV} and we have added two more at low
densities. In the following, we identify these densities by a
label from 1 (high density close to the crust-core transition) to
12 (low density below neutron drip) and will refer to twelve
density regions. In Fig. \ref{dnddx} and in Fig.~\ref{ZNRwXFig} we
plot some of the properties of the Wigner-Seitz cells as a
function of density. In particular, in Fig. \ref{dnddx} we show
  the neutron density at the cell center and the cell border, the cluster proton
  fraction at the cluster center and the neutron skin thickness
  ${\Theta=R_n-R_p}$, with ${R_i^2=5/3<r_i^2>}$,
 and in Fig.~\ref{ZNRwXFig},
 we plot the Wigner-Seitz proton number $Z$, neutron number
$N$, radius $R_{ws}$, total proton fraction $x$, droplet number
$A$ and droplet proton fraction $Z/A$. The droplet nucleon number
$A$ has been estimated from the radius $R_{ws}$ and the neutron
density at $R_{ws}$, namely, ${A=Z+
N[1-V^i(R_{ws})\rho_N(R_{ws})]}$. Whenever available data from
\cite{NV,grill,Haensel} have been included for comparison. In
order to better understand the behaviour of the NL3 interaction we
have added some points at high density (regions from 1 to 3).

The main differences occur above $\rho=0.02-0.05$ fm$^{-3}$. Below
this density, except for the IU-FSU and NL3, all TF results lie
between an upper and lower bound defined by the HFB and HF
calculations. Above $\rho=0.05$ fm$^{-3}$ non-spherical geometries may arise
with lower energy within the TF calculation, and this explains
partially the differences. The CLD calculation for the SLy4
interaction of~\cite{Haensel} also follows the main trend obtained
with the relativistic models.

We will first discuss the proton number as a function of density.
Different properties of the models explain the existing
differences in the results.
NL3 and IU-FSU have, respectively, the smallest and largest proton
number. This is a clear effect of the slope $L$, the large value
of NL3 and the small value of IU-FSU, and corresponding surface
tensions: in IU-FSU neutrons do not drip out so easily, the
central droplet density is larger, and according to the liquid
droplet model~\cite{pethick95}, the proton number increases with
the surface energy; the opposite occurs with NL3.
 NL3$_{\omega\rho}$ only differs from NL3 in the isospin channel: a
larger $L$ clearly gives a larger proton number. The interaction
DD-ME$\delta$ has, next to NL3, the smallest proton number and
Wigner-Seitz radius in a large part of the density range
considered, due to a smaller surface energy at large proton
fraction which favors smaller droplets. FSU has a larger effective
nucleon mass at saturation than all the other models, except
IU-FSU. In \cite{estal99} it was shown that a larger effective
mass favors a smaller surface thickness and, therefore, a larger
nucleon number inside the droplet is expected, including  a larger
proton number.

The results reported in Table~\ref{TabZNRw} show that in the TF
calculations the proton number does not change considerably close
to the transition to the outer crust and lies between 39 and 46.
The proton number found with the HFB calculations converges to
values around $Z=36$, which is the proton number
that characterizes the nuclei at the drip density
(e.g.~\cite{RocaOut}). This shell effect can not be reproduced by
the TF calculations. The proton number obtained using the DD-ME2,
DD-ME$\delta$, NL3$_{\omega\rho}$ and FSU show a similar behaviour
to the one found in~\cite{Haensel}, where shell effects are also
neglected: below $\rho\sim 10^{-3}\mbox{fm}^{-3}$ all these models
predict a larger proton number than the HFB~\cite{grill} or the
HF~\cite{NV} calculations. This results are also similar to the
ones calculated in~\cite{rbp} within the liquid drop model or
in~\cite{oyamastu1} within a TF calculation with energy density
functionals constructed in order to reproduce nuclear matter
properties. In this last work it is seen that $Z$ is sensitive to
the properties of the EOS. On the other hand in~\cite{oyamatsu2},
a relativistic Brueckner-Hartree-Fock calculation was performed
and quite small proton numbers were obtained, very similar to NL3.
Our results also agree with the conclusions of~\cite{oyamatsu07}
where, within a Thomas Fermi calculation applied to a macroscopic
nuclear model, it was shown that the proton number is larger for
smaller values of $L$.

In regions 1 and 2 the proton number found with the TF
calculations has a consistent drop (Fig.~\ref{ZNRwXFig}). This is a
behaviour also occurring in other works (e.g.~\cite{NV},
\cite{baldo}, \cite{oyamatsu2} and \cite{Haensel}) and it is
related with both the decrease of the Wigner-Seitz radius and the
increase of the volume occupied with nuclear matter.

The neutron numbers obtained with all interactions are very
similar below $\rho\sim 5 \times 10^{-3}$ fm$^{-3}$ and they are
comparable to the ones calculated within the HFB \cite{grill} and
HF \cite{NV} calculations where shell effects are included, (see
Table~\ref{TabZNRw} and Fig.~\ref{ZNRwXFig}). Above $\rho\sim 5
\times 10^{-3}$  fm$^{-3}$ model properties explain the existing
differences.

NL3 and DD-ME$\delta$ interaction give neutron numbers smaller
than the other interactions { (except for NL3 in region 2)}. The
neutron numbers found with NL3$_{\omega\rho}$ (in the regions 1
and 2) and IU-FSU (in the region 1) are  higher than what has been
found with the other interactions. In fact, IU-FSU predicts the
rod shape in region 1. If the droplet shape would have been
considered in this region for IU-FSU the cells would have been
characterized by $Z=102$, $N=2066$ and $R_{ws}=22.2$ fm, much
bigger than what has been found with all the other interactions.
As a whole the neutron number calculated in all the considered
interactions is coherent with the results found in the literature.
It increases consistently with the density and only at very high
density does it decreases, as found in~\cite{Haensel},
\cite{grill} and \cite{NV}. We conclude that, as expected, the
neutron number is not strongly linked to the shell effect.

The TF and HFB neutron numbers are very similar just in the middle
density range (corresponding to regions 5, 6 and 7). For larger
densities (regions 2, 3 and 4) the neutron numbers found with the
HFB calculations are lower. Only in the region 3 the neutron
numbers are similar, but this is a consequence of a big difference
on the proton number. Finally, as expected, an important
difference with respect to~\cite{grill} occurs in the low density
regions. As also found with the proton number, the HFB results
converge to the closed shell neutron number ($N=82$) which
characterizes the neutron drip density and the initial part of the
outer crust (e.g.~\cite{RocaOut}), while the TF calculations,
which does not include shell effects, cannot reproduce this
behaviour. Furthermore, looking at the neutron chemical potential
(Tab.~\ref{TabEMuX}), we conclude that the neutron drip occurs
above region 10 in the HFB calculations, due to the shell effects,
while the TF calculations predict this transition in the region
12, which is characterized by quite low density. {However, with
the NL3 parametrization we have found a drip density similar to
that found with the HFB calculations}. TF results are very similar
to those found by Negele and Vautherin~\cite{NV}: only at the very
high density (regions 1 and 2) there is a difference and the
neutron number found in~\cite{NV} is much larger than that found
with TF.

The cell neutron number is closely linked to the radius of the
Wigner-Seitz cells (Tab.~\ref{TabZNRw} and Fig.~\ref{ZNRwXFig}).
The cell radius is quite independent on the nuclear interaction
and the qualitative behaviour obtained in the TF calculation is
equal to what has been found with HFB \cite{grill} and HF
\cite{NV}. Some quantitative differences identified are: a) the
cells calculated with NL3 and DD-ME$\delta$ interaction are
slightly smaller than those found with all the other interactions,
which is coherent with the smaller neutron number predicted by
these interactions (only region 2 is out of this trend in NL3); b)
as with the neutron number the cell radius obtained using the TF
and HFB approach are very similar in the middle density regions
(from 5 to 7) but differ in the low and high density ranges, where
the HFB radius is smaller.

In region 1, the lowest energy cell shape predicted by IU-FSU is
the rod and, as it will be described in the next section, a change
in the cell shape is always correlated with a drop in the value of
the cell radius (Fig.~\ref{ZNRwFigPat}). However, this does not
occur in this case because IU-FSU predicts values higher than
those found with all the other interactions.

The proton fraction calculated with all interactions is very
similar (Tabs.~\ref{TabEMuX}, \ref{TabMuXPat} and
Fig.~\ref{ZNRwXFig}). Only the values calculated with the IU-FSU
and NL3 interactions are different from the general trend at
higher densities (Tab.~\ref{TabEMuX} in the regions from 1 to 4
and Tab.~\ref{TabMuXPat}). The behaviour of the proton fraction is
similar to those found in the
literature~\cite{Sidney,Gogelein,NV,oyamatsu07}: its value
decreases with increasing the density until, at high densities
close to the crust-core transition, it changes its slope in order
to reproduce the correct behaviour of homogeneous matter
(Fig.~\ref{ZNRwXFig}). IU-FSU has the largest symmetry energy at
$\rho\sim 0.5\rho_0$ which favors isospin symmetry, NL3 has the
smallest symmetry energy which favors a smaller proton fraction.
In general a larger symmetry energy favors matter with a smaller
isospin asymmetry. This is true for homogeneous matter, but the
discussion of non-homogenous matter requires care: properties of
clusters are closely dependent on the surface energy which defines
how favorable is the creation of clusters.

The proton fractions obtained
in~\cite{Sidney} and \cite{Gogelein} for the RMF model are smaller
than the ones obtained in the present model, except for the NL3
parametrization. This is mainly due to the characteristics of the
RMF used: both of them have a large symmetry energy slope at
saturation and therefore, a smaller symmetry energy at
subsaturation densities, which favors large isospin asymmetries.
Similar results have been obtained in~\cite{Maruyama}. The
calculated proton fraction in~\cite{Gogelein} with the SLy4
parametrization is consistent both with the CLD model and the HFB
calculation, which have used the same parametrization, and are
just slightly larger than the ones obtained with all the RMF
models discussed in the present work except IU-FSU.
In~\cite{Gogelein} it was also shown that the shell effects were
not very important and its main effect was to determine a slightly
larger proton fraction.

Some properties of the clusters
reflect clearly the density dependence of the symmetry energy
(see Fig.~\ref{dnddx} and \ref{ZNRwXFig}).  The cluster nucleon number
above $\rho=5\times 10^{-3}$~fm$^{-3}$ is correlated with
$L$ while below that density it is not very sensitive to the
model. The nucleon number of NL3 clusters decreases above ${10^{-2}\ fm^{-3}}$ and at the
crust-core transition is less than 100. On the other hand IU-FSU and SLy4, the models with the
smallest slopes $L$, predict clusters with more than 500 nucleons
close to the crust-core transition. All the other models predict
clusters that do not go beyond 200-250 nucleons.

The cluster proton fraction decreases, as expected, with density
and estabilizes around 0.25 for models with $L=51-60$ MeV. For
both models with the smallest $L$, IU-FSU and SLy4 it decreases
continuously reaching proton fractions below 0.15 while NL3 just
before the crust-core transition has a strong increase of Z/A. We
note, however, that as expected from the symmetry energy close to
saturation, NL3, with the largest symmetry energy,  has the
largest proton fraction at the center of the cluster, while IU-FSU
has the smallest. The droplet overall proton fraction is then
defined by the clusters neutron skin (Fig.~\ref{dnddx}), which is
much larger in NL3.

\subsection{Pasta phase regions}\label{pasta}

The properties of the pasta phase regions of the inner crust are
reported in the tables from \ref{TabZNPat} to \ref{TabTra} and in
Fig.~\ref{ZNRwFigPat} for the relativistic models.

In the pasta phase regions the matter is studied considering the
three different cell shapes: the droplet, the rod and the slab.
The transition densities between these different shapes are
reported in table~\ref{TabTra} for all interactions. In this table
we have also included the results of Ref.~\cite{Gogelein}. All the
three shapes appear in the inner crust except for NL3, which only
predicts droplets. These results agree with those of
Refs.~\cite{Sidney} and \cite{Sidney3}, where, using the TF
approach and the NL3 parametrization, it was predicted that in
$\beta-$equilibrium condition only the droplet cell shape would
appear. Droplets were also the only configuration obtained in
Refs.~\cite{Haensel} and \cite{Maruyama}. However, contrary
to~\cite{Haensel} where the CLD was applied, in~\cite{Gogelein}
all the three shapes were obtained with SLy4 using different
frameworks, both TF and HF. In~\cite{oyamatsu07} it was shown that
models with a large $L$ would not predict the existence of pasta
shapes in $\beta$-equilibrium matter. Thus we expect that the RMF
parametrization used in~\cite{Maruyama} has a large slope $L$. The
IU-FSU shape transitions occur at densities similar to those
obtained in~\cite{Gogelein} for SLy4 within the HF calculation.
These two models predict similar values of L.

The crust-core transition densities found in this work, and
indicated in Table~\ref{TabTra}, agree with the ones found in the
literature within a dynamical spinodal calculation: SLy4 at
${\rho_t=0.080\ \mbox{fm}^{-3}}$, NL3 at ${\rho_t=0.054\
\mbox{fm}^{-3}}$, DD-ME2 at ${\rho_t=0.072\ \mbox{fm}^{-3}}$, FSU
at ${\rho_t=0.074\ \mbox{fm}^{-3}}$~\cite{Isaac},
NL3$_{\omega\rho}$ at ${\rho_t=0.0855\
\mbox{fm}^{-3}}$~\cite{Pais} and IU-FSU at ${\rho_t=0.087\
\mbox{fm}^{-3}}$~\cite{iufsu}. As expected results from the pasta
calculation are just slightly larger. The crust core transition
density is expected to occur within the metastable region between
the spinodal and the binodal surfaces. However, for very
asymmetric matter as $\beta$-equilibrium matter, this region is
almost nonexistent, since the binodal and the spinodal surfaces
are very close. These results confirm the conclusions of
reference~\cite{Sidney3}, namely that the estimation of the crust
core transition from the dynamical spinodal is a good one.

The proton and neutron numbers as well as the Wigner-Seitz radius
in the pasta phase are reported in the tables~\ref{TabZNPat} and
\ref{TabRWEPat} and in Fig.~\ref{ZNRwFigPat}. Comparing the
behavior of the different models we conclude that: a) the NL3 and
IU-FSU interactions are quite different from all the others with a
much smaller or larger $Z$, $N$ and $R_{ws}$. The very different
surface energy obtained within these two models explains this
difference; b) all the other models behave in a very similar way,
and differences may be explained comparing the symmetry energy
slopes  within the models built using the same formalism,
namely NLWM or density dependent hadronic models. A smaller $L$
favors larger $Z$, $N$, and $R_{ws}$ in NL3$_{\omega\rho}$ with
respect to FSU. A smaller effective mass and incompressibility in
DD-ME$\delta$ favors a smaller surface energy and therefore,
smaller $Z$, $N$, and $R_{ws}$ than in DD-ME2; c) the behavior of
the surface energy with the proton asymmetry explains small
differences between models. Close to the crust-core transition
asymmetries are large and models with larger surface energies will
suffer shape transitions at larger densities for similar cell
sizes.

At the shape transitions both the protons and neutron numbers have
a sudden change because these numbers have a strong dependence on
the cell volume~(\ref{VolR}) and (\ref{VolS}) and thus on the
parameter $l$, which has been arbitrary chosen. However this
parameter has just an influence on the proton and neutron number
values, it cannot change their behaviour or the calculated matter
properties, namely their dependence on the density, the proton
fraction, the energy per baryon or the chemical potentials. The
cell radius decreases until densities close to a shape transition
where it stabilizes or a small increase occurs just before the
transition. The shape transition is then characterized by a
decrease of cell size. Similar behavior was described
in~\cite{Sidney,Maruyama}.

A comparison with the results already published in the literature
is not always possible. Indeed in~\cite{grill} just the standard
inner crust has been studied, while the results found in~\cite{NV}
at very high density are not realistic because the droplet cell
shape was imposed. In~\cite{Sidney,Gogelein,Maruyama} consistent
calculations have been performed, but a detailed description of
the cell structure ($N$,$Z$ and $R_{ws}$) was not reported. {The
cell structure has been reported in~\cite{Haensel} (Tab. 1 and 2
of Ref.~\cite{Haensel}) where it is possible to see that the
proton {and neutron numbers} have a behaviour similar to what has
been found with our TF calculations: after an initial increase
until quite high densities,  then they decrease and finally they
grow again until the crust-core transition (${\rho=7.5959\times
10^{-2}\ \mbox{fm}^{-3}}$). Also the cell radius has a behaviour
very similar to the one obtained in this work: it monotonically
decreases until densities close to the crust-core transition
(${\rho=7.0154\times 10^{-2}\ \mbox{fm}^{-3}}$), then it grows
until the transition.

One interesting conclusion is that, except for NL3, all the models
studied predict  slab like configurations in $\beta$-equilibrium
matter. According to~\cite{luc2011} low energy collective modes
with an important contribution to the specific heat could be
excited in these 'lasagna'-like pastas.

The density in which non-spherical shapes appear was discussed
in~\cite{pethick95}, and an estimation of a nuclear filling
fraction of 1/8 was obtained. If no  dripped neutrons occur this
fraction translates into a average cell density of $n_i/8$, where
$n_i$ is the central density of the droplet. However, this density
will be larger if the dripped-neutron density is non zero. In
fact, we confirm that the models with the lowest rod shape onset
have the smallest neutron density at the cell border. This density
is closely correlated with  the slope $L$, with a smaller $L$
corresponding to a smaller neutron density. NL3 is an  exception
because the crust-core transition occurs below this limit.
 The onset of the slab geometry between different
models follows a behavior similar to the rod onset.

\section{Conclusions}

We have studied the inner crust properties of neutron stars within
a self-consistent Thomas Fermi approach developed in
\cite{Sidney,Sidney2} for relativistic nuclear models. Several
relativistic nuclear models have been used both with non linear
meson terms and constant couplings, and with density dependent
coupling constants. The results have been compared with
calculations obtained within the HFB, and the
HF~\cite{NV,grill,Gogelein} formalisms, with the compressible
liquid drop model \cite{Haensel} and with a macroscopic nuclear
model \cite{oyamatsu07}.

It has been shown that the main properties of the Wigner-Seitz
cells obtained within the HFB and HF formalisms are reproduced,
namely, the average proton number and the neutron number and the
Wigner-Seitz cell radius. As expected,  proton shell effects are
missing.

The properties of the models used are reflected on the cluster
structure. It was seen that a small symmetry energy slope $L$ gave
rise to larger cells, with a larger proton and neutron number,
while the opposite occurs for a large $L$. Models with a similar
symmetry energy ($\sim31-32$ MeV) and  slope $L$ ($\sim 50-60$) at
saturation density were shown to behave in a similar way, both in
the droplet phase and the pasta phase regions. On the other hand
models  like NL3, with a very large symmetry energy and slope $L$
and  IU-FSU, with a quite small $L$, have shown quite different
behaviors. NL3 did not present any pasta phases in the inner crust
of $\beta$-equilibrium matter, and predicted the smallest proton
and neutron numbers, and Wigner-Seitz radius in almost all the
inner crust range of densities.
 On the other hand, IU-FSU predicts
a quite low density for the onset of the pasta phase, where all
the other models still predict the existence of droplets. The
occurrence of the slab shape  occurs at a lower densities than in
all the other models. However, the IU-FSU crust-core transition
density is the largest one and  above 0.01 fm$^{-3}$  IU-FSU
presents the largest clusters with more than the double of
nucleons. All the models, except NL3,
 predict the existence of  slab like configurations in $\beta$-equilibrium
matter. These 'lasagna'-like structures may have an important contribution to
the specific heat of the crust ~\cite{luc2011}.

The  size and composition of the clusters will have an important
effect on the transport properties of the crust. In
\cite{sonoda2007} it was shown how the pasta structures could
affect the neutrino transport, namely a more uniform distribution
of matter, as occurs in NL3 with a larger neutron drip, or a
larger range with non-spherical pasta structures could reduce the
cross section for elastic neutrino scattering from pasta phases
via weak neutral current, and therefore the neutrino opacity.

The effect of the pairing correlations, which were missing in the
present work, on the inner crust clusters and the effect of the
size and composition of the clusters on the transport properties
of the crust will  be investigated.

\section*{Acknowledgments}
This work has been partially supported by QREN/FEDER, the
Programme COMPETE, and FCT (Portugal) under the
projects PTDC/FIS/113292/2009
 and CERN/FP/116366/2010, by the Capes/FCT n. 232/09 bilateral
collaboration  and by COMPSTAR, an ESF Research Networking
Programme.

\begin{table*}
  \centering
  \begin{tabular}{l|cccccccccccccc}
    \hline\hline
     && $\rho_0\ [\mbox{fm}^{-3}]$ && $E_0\ [\mbox{MeV}]$ && $K_0\ [\mbox{MeV]}$ && $E_{sym}\ [\mbox{MeV}]$ && $L\ [\mbox{MeV}]$ && $M^*/M$ && $m_\sigma\ [MeV]$\\
    \hline
    SLy4               && $0.159$ && $-15.97$ && $229.8$ && $31.8$ && $45.3$ && $0.695$ && $    -$\\
    NL3                && $0.148$ && $-16.24$ && $270.7$ && $37.3$ &&$118.3$ && $0.600$ && $508.2$\\
    DD-ME2             && $0.152$ && $-16.14$ && $250.8$ && $32.3$ && $51.4$ && $0.609$ && $550.0$\\
    DD-ME$\delta$      && $0.152$ && $-16.12$ && $219.1$ && $32.4$ && $52.9$ && $0.572$ && $566.2$\\
    NL3$_{\omega\rho}$ && $0.148$ && $-16.30$ && $272.0$ && $31.7$ && $55.2$ && $0.600$ && $508.2$\\
    FSU                && $0.148$ && $-16.30$ && $230.0$ && $32.6$ && $60.5$ && $0.620$ && $491.5$\\
    IU-FSU             && $0.155$ && $-16.40$ && $231.2$ && $31.3$ && $47.2$ && $0.620$ && $491.5$\\
    \hline\hline
  \end{tabular}
  \caption{Nuclear matter properties at the saturation density (density,
    binding energy, incompressibility, symmetry energy, symmetry energy slope
    and effective mass) and the $\sigma$ meson mass.}\label{TabSat}
\end{table*}

\begin{table*}
  \centering
\begin{tabular}{l|cccccccccccccccccccccccccc}
  \hline\hline
  \multicolumn{1}{c|}{Regions} && && 1 && 2 && 3 && 4 && 5 && 6 && 7 && 8 && 9 && 10 && 11 && 12 \\
  \multicolumn{1}{c|}{$\rho\ [10^{-3}\mbox{fm}^{-3}]$} && && $47.1$ && $20.2$ && $8.83$ && $5.72$ && $3.70$ && $1.58$ && $0.871$ && $0.595$ && $0.396$ && $0.276$ && $0.259$ && $0.188$ \\
  \hline
  \multicolumn{1}{c|}{$Z$} &&&&&&&&&&&&&&&&&&&&&&&&&&\\
HFB                && && $ -$ && $40$ && $54$ && $40$ && $46$ && $50$ && $50$ && $36$ && $38$ && $36$ && $38$ && $38$ \\
N$\&$V             && && $40$ && $50$ && $50$ && $50$ && $50$ && $40$ && $40$ && $40$ && $40$ && $40$ && $ -$ && $ -$ \\
NL3                && && $16$ && $27$ && $32$ && $34$ && $36$ && $38$ && $38$ && $39$ && $39$ && $39$ && $39$ && $39$ \\
DD-ME2             && && $33$ && $39$ && $42$ && $43$ && $43$ && $43$ && $43$ && $43$ && $43$ && $42$ && $43$ && $42$ \\
DD-ME$\delta$      && && $32$ && $39$ && $42$ && $42$ && $42$ && $42$ && $41$ && $41$ && $41$ && $41$ && $41$ && $40$ \\
NL3$_{\omega\rho}$ && && $39$ && $42$ && $44$ && $44$ && $44$ && $44$ && $44$ && $44$ && $43$ && $43$ && $43$ && $42$ \\
FSU                && && $37$ && $44$ && $46$ && $46$ && $46$ && $46$ && $46$ && $46$ && $45$ && $45$ && $45$ && $44$ \\
IU-FSU             && && $82^r$ && $63$ && $54$ && $52$ && $50$ && $48$ && $47$ && $47$ && $47$ && $46$ && $46$ && $45$\\
  \multicolumn{1}{c|}{$N$} &&&&&&&&&&&&&&&&&&&&&&&&&&\\
HFB                && && $   -$ && $1018$ && $1324$ && $ 732$ && $ 740$ && $ 454$ && $ 316$ && $ 174$ && $ 120$ && $  82$ && $  82$ && $  82$ \\
N$\&$V             && && $1460$ && $1750$ && $1300$ && $1050$ && $ 900$ && $ 460$ && $ 280$ && $ 210$ && $ 160$ && $ 140$ && $   -$ && $   -$ \\
NL3                && && $1019$ && $1816$ && $1202$ && $ 925$ && $ 718$ && $ 410$ && $ 259$ && $ 193$ && $ 134$ && $  98$ && $  96$ && $  89$ \\
DD-ME2              && && $1116$ && $1463$ && $1213$ && $1016$ && $ 808$ && $ 482$ && $ 313$ && $ 229$ && $ 160$ && $ 110$ && $ 106$ && $  92$ \\
DD-ME$\delta$      && && $1107$ && $1301$ && $1087$ && $ 908$ && $ 739$ && $ 448$ && $ 286$ && $ 211$ && $ 147$ && $ 105$ && $  99$ && $  86$ \\
NL3$_{\omega\rho}$ && && $1300$ && $1590$ && $1291$ && $1045$ && $ 821$ && $ 476$ && $ 308$ && $ 224$ && $ 152$ && $ 108$ && $ 102$ && $  92$ \\
FSU                && && $1192$ && $1482$ && $1243$ && $1035$ && $ 835$ && $ 503$ && $ 328$ && $ 241$ && $ 164$ && $ 116$ && $ 110$ && $  96$ \\
IU-FSU             && && $1655^r$ && $1417$ && $1188$ && $1031$ && $ 840$ && $ 510$ && $ 332$ && $ 247$ && $ 173$ && $ 120$ && $ 113$ && $  95$ \\
 \multicolumn{1}{c|}{$R_{ws}\ [\mbox{fm}]$} &&&&&&&&&&&&&&&&&&&&&&&&&&\\
HFB                && && $   -$ && $23.2$ && $33.4$ && $31.8$ && $37.0$ && $42.4$ && $46.4$ && $43.8$ && $45.6$ && $46.8$ && $48.0$ && $53.4$ \\
N$\&$V             && && $19.6$ && $27.6$ && $33.1$ && $35.7$ && $39.3$ && $42.2$ && $44.3$ && $46.3$ && $49.2$ && $53.6$ && $   -$ && $   -$ \\
NL3                && && $17.4$ && $27.9$ && $32.2$ && $34.2$ && $36.5$ && $40.8$ && $43.3$ && $45.3$ && $47.1$ && $49.1$ && $49.9$ && $54.6$ \\
DD-ME2              && && $18.0$ && $26.1$ && $32.4$ && $35.4$ && $38.0$ && $43.0$ && $46.0$ && $47.8$ && $49.6$ && $50.8$ && $51.6$ && $55.5$ \\
DD-ME$\delta$      && && $17.9$ && $25.1$ && $31.3$ && $34.1$ && $36.9$ && $42.0$ && $44.8$ && $46.6$ && $48.4$ && $50.1$ && $50.5$ && $54.3$ \\
NL3$_{\omega\rho}$ && && $18.9$ && $26.8$ && $33.0$ && $35.7$ && $38.2$ && $42.9$ && $45.9$ && $47.6$ && $49.0$ && $50.7$ && $51.1$ && $55.5$ \\
FSU                && && $18.4$ && $26.2$ && $32.7$ && $35.6$ && $38.5$ && $43.7$ && $46.8$ && $48.7$ && $50.1$ && $51.8$ && $52.3$ && $56.3$ \\
IU-FSU             && && $19.8^r$ && $26.0$ && $32.3$ && $35.6$ && $38.6$ && $43.9$ && $47.0$ && $49.1$ && $51.0$ && $52.3$ && $52.7$ && $56.3$ \\
  \hline\hline
\end{tabular}
\caption{Cell proton number, neutron number and Wigner-Seitz
radius for different nuclear interactions at the considered
densities. HFB and N$\&$V refer respectively to the results found
in~\cite{grill} and~\cite{NV}. In the region 1, for the IU-FSU
interaction, the cell shape is the rod.}\label{TabZNRw}
\end{table*}

\begin{table*}
  \centering
\begin{tabular}{l|ccccccccccccc}
  \hline\hline
  \multicolumn{1}{c|}{Regions} & & 1 & 2 & 3 & 4 & 5 & 6 & 7 & 8 & 9 & 10 & 11 & 12 \\
  \multicolumn{1}{c|}{$\rho\ [10^{-3}\mbox{fm}^{-3}]$} & & $47.1$ & $20.2$ & $8.83$ & $5.72$ & $3.70$ & $1.58$ & $0.871$ & $0.595$ & $0.396$ & $0.276$ & $0.259$ & $0.188$ \\
  \hline
  \multicolumn{1}{c|}{$E/A\ [MeV]$} &&&&&&&&&&&&&\\
HFB                & & $     -$ & $ 4.747$ & $ 3.025$ & $ 2.283$ & $ 1.645$ & $ 0.612$ & $-0.051$ & $-0.490$ & $-1.063$ & $-1.691$ & $-1.830$ & $-2.441$ \\
N$\&$V             & & $ 6.428$ & $ 4.097$ & $ 2.610$ & $ 1.996$ & $ 1.465$ & $ 0.541$ & $-0.050$ & $-0.462$ & $-0.962$ & $-1.425$ & $     -$ & $     -$ \\
NL3                & & $ 4.491$ & $ 3.054$ & $ 2.438$ & $ 2.096$ & $ 1.735$ & $ 0.993$ & $ 0.422$ & $-0.004$ & $-0.545$ & $-1.130$ & $-1.244$ & $-1.801$ \\
DD-ME2             & & $ 6.847$ & $ 4.494$ & $ 3.130$ & $ 2.560$ & $ 2.051$ & $ 1.174$ & $ 0.584$ & $ 0.170$ & $-0.341$ & $-0.890$ & $-1.002$ & $-1.581$ \\
DD-ME$\delta$      & & $ 6.986$ & $ 4.796$ & $ 3.313$ & $ 2.681$ & $ 2.123$ & $ 1.186$ & $ 0.569$ & $ 0.139$ & $-0.391$ & $-0.959$ & $-1.075$ & $-1.666$ \\
NL3$_{\omega\rho}$ & & $ 6.750$ & $ 4.354$ & $ 3.020$ & $ 2.474$ & $ 1.984$ & $ 1.122$ & $ 0.526$ & $ 0.100$ & $-0.430$ & $-1.003$ & $-1.119$ & $-1.702$ \\
FSU                & & $ 6.987$ & $ 4.660$ & $ 3.202$ & $ 2.601$ & $ 2.072$ & $ 1.173$ & $ 0.575$ & $ 0.155$ & $-0.362$ & $-0.917$ & $-1.030$ & $-1.609$ \\
IU-FSU             & & $ 8.531$ & $ 5.630$ & $ 3.645$ & $ 2.884$ & $ 2.252$ & $ 1.250$ & $ 0.623$ & $ 0.194$ & $-0.325$ & $-0.881$ & $-0.994$ & $-1.587$ \\
  \multicolumn{1}{c|}{$\mu_N\ [MeV]$} &&&&&&&&&&&&&\\
HFB                & & $     -$ & $ 7.178$ & $ 4.600$ & $ 3.713$ & $ 3.117$ & $ 1.730$ & $ 1.036$ & $ 0.724$ & $ 0.327$ & $-2.660$ & $-3.278$ & $-3.287$ \\
N$\&$V             & & $10.900$ & $ 6.500$ & $ 4.200$ & $ 3.300$ & $ 2.600$ & $ 1.400$ & $ 1.000$ & $ 0.600$ & $ 0.300$ & $ 0.200$ & $     -$ & $     -$ \\
NL3                & & $ 7.285$ & $ 3.242$ & $ 2.550$ & $ 2.261$ & $ 1.932$ & $ 1.253$ & $ 0.816$ & $ 0.556$ & $ 0.288$ & $-0.031$ & $-0.189$ & $-0.809$ \\
DD-ME2             & & $10.118$ & $ 5.842$ & $ 3.883$ & $ 3.147$ & $ 2.511$ & $ 1.509$ & $ 0.971$ & $ 0.677$ & $ 0.398$ & $ 0.145$ & $ 0.094$ & $-0.442$ \\
DD-ME$\delta$      & & $ 9.717$ & $ 6.218$ & $ 4.224$ & $ 3.394$ & $ 2.679$ & $ 1.572$ & $ 0.995$ & $ 0.689$ & $ 0.397$ & $ 0.144$ & $ 0.094$ & $-0.515$ \\
NL3$_{\omega\rho}$ & & $10.050$ & $ 5.712$ & $ 3.715$ & $ 3.009$ & $ 2.409$ & $ 1.455$ & $ 0.935$ & $ 0.646$ & $ 0.365$ & $ 0.105$ & $ 0.046$ & $-0.568$ \\
FSU                & & $ 9.938$ & $ 6.130$ & $ 4.040$ & $ 3.244$ & $ 2.569$ & $ 1.526$ & $ 0.974$ & $ 0.677$ & $ 0.390$ & $ 0.133$ & $ 0.086$ & $-0.526$ \\
IU-FSU             & & $11.659$ & $ 7.907$ & $ 4.933$ & $ 3.822$ & $ 2.934$ & $ 1.671$ & $ 1.051$ & $ 0.731$ & $ 0.432$ & $ 0.179$ & $ 0.131$ & $-0.425$ \\
 \multicolumn{1}{c|}{$x$} &&&&&&&&&&&&&\\
HFB                & & $     -$ & $ 0.038$ & $ 0.039$ & $ 0.052$ & $ 0.059$ & $ 0.099$ & $ 0.137$ & $ 0.171$ & $ 0.241$ & $ 0.305$ & $ 0.317$ & $ 0.317$ \\
N$\&$V             & & $ 0.027$ & $ 0.028$ & $ 0.037$ & $ 0.045$ & $ 0.053$ & $ 0.080$ & $ 0.125$ & $ 0.160$ & $ 0.200$ & $ 0.222$ & $     -$ & $     -$ \\
NL3                & & $ 0.015$ & $ 0.015$ & $ 0.026$ & $ 0.035$ & $ 0.048$ & $ 0.085$ & $ 0.128$ & $ 0.168$ & $ 0.225$ & $ 0.285$ & $ 0.289$ & $ 0.305$ \\
DD-ME2             & & $ 0.029$ & $ 0.026$ & $ 0.033$ & $ 0.041$ & $ 0.051$ & $ 0.082$ & $ 0.121$ & $ 0.158$ & $ 0.212$ & $ 0.276$ & $ 0.289$ & $ 0.313$ \\
DD-ME$\delta$      & & $ 0.028$ & $ 0.029$ & $ 0.037$ & $ 0.044$ & $ 0.054$ & $ 0.086$ & $ 0.125$ & $ 0.163$ & $ 0.218$ & $ 0.281$ & $ 0.293$ & $ 0.317$ \\
NL3$_{\omega\rho}$ & & $ 0.029$ & $ 0.026$ & $ 0.033$ & $ 0.040$ & $ 0.051$ & $ 0.085$ & $ 0.125$ & $ 0.164$ & $ 0.221$ & $ 0.285$ & $ 0.297$ & $ 0.313$ \\
FSU                & & $ 0.030$ & $ 0.029$ & $ 0.036$ & $ 0.043$ & $ 0.052$ & $ 0.084$ & $ 0.123$ & $ 0.160$ & $ 0.215$ & $ 0.280$ & $ 0.290$ & $ 0.314$ \\
IU-FSU             & & $ 0.047$ & $ 0.043$ & $ 0.043$ & $ 0.048$ & $ 0.056$ & $ 0.086$ & $ 0.124$ & $ 0.160$ & $ 0.214$ & $ 0.277$ & $ 0.289$ & $ 0.321$ \\
  \hline\hline
\end{tabular}
\caption{Cell energy per baryon, neutron chemical potential and
proton fraction for several nuclear interactions at the considered
densities. HFB and N$\&$V refer respectively to the results found
in~\cite{grill} and~\cite{NV}. In the region 1, for the IU-FSU
interaction, the cell shape is the rod.}\label{TabEMuX}
\end{table*}

\begin{table*}
  \centering
\begin{tabular}{c|cccccccccccccc}
  \hline\hline
 $\rho$ & &\multicolumn{6}{c}{Z} & &\multicolumn{6}{c}{N} \\
\cline{3-8}\cline{10-15}
   $[\mbox{fm}^{-3}]$& & NL3 & DD-ME2 & DD-ME$\delta$ & NL3$_{\omega\rho}$ & FSU & IU-FSU & &
   NL3 & DD-ME2 & DD-ME$\delta$ & NL3$_{\omega\rho}$ & FSU & IU-FSU \\
  \hline
$8.94\times 10^{-2}$ & &       &       &       &   $-$ &       &   $-$ & &        &        &        &    $-$ &        &    $-$ \\\cline{8-8}\cline{15-15}
$8.88\times 10^{-2}$ & &       &       &       &   $-$ &       & $173$ & &        &        &        &    $-$ &        & $3594$ \\
$8.35\times 10^{-2}$ & &       &       &       &   $-$ &       & $117$ & &        &        &        &    $-$ &        & $2418$ \\\cline{6-6}\cline{13-13}
$8.29\times 10^{-2}$ & &       &       &       &  $67$ &       & $114$ & &        &        &        & $1680$ &        & $2355$ \\
$7.82\times 10^{-2}$ & & $  -$ & $  -$ & $  -$ & $ 60$ & $  -$ & $101$ & & $   -$ & $   -$ & $   -$ & $1557$ & $   -$ & $2071$ \\
$7.66\times 10^{-2}$ & & $  -$ & $  -$ & $  -$ & $ 59$ & $  -$ & $ 99$ & & $   -$ & $   -$ & $   -$ & $1548$ & $   -$ & $2027$ \\\cline{5-5}\cline{12-12}
$7.51\times 10^{-2}$ & & $  -$ & $  -$ & $ 49$ & $ 59$ & $  -$ & $ 97$ & & $   -$ & $   -$ & $1396$ & $1565$ & $   -$ & $1981$ \\\cline{6-6}\cline{7-7}\cline{13-13}\cline{14-14}
$7.35\times 10^{-2}$ & & $  -$ & $  -$ & $ 48$ & $ 49$ & $ 56$ & $ 95$ & & $   -$ & $   -$ & $1385$ & $1319$ & $1503$ & $1935$ \\\cline{4-4}\cline{11-11}
$7.20\times 10^{-2}$ & & $  -$ & $ 54$ & $ 47$ & $ 47$ & $ 53$ & $ 94$ & & $   -$ & $1467$ & $1373$ & $1280$ & $1439$ & $1910$ \\\cline{5-5}\cline{12-12}
$7.04\times 10^{-2}$ & & $  -$ & $ 51$ & $ 37$ & $ 46$ & $ 51$ & $ 93$ & & $   -$ & $1403$ & $1095$ & $1267$ & $1401$ & $1885$ \\
$6.88\times 10^{-2}$ & & $  -$ & $ 50$ & $ 36$ & $ 45$ & $ 50$ & $ 92$ & & $   -$ & $1394$ & $1079$ & $1254$ & $1391$ & $1861$ \\\cline{4-4}\cline{11-11}
$6.73\times 10^{-2}$ & & $  -$ & $ 41$ & $ 35$ & $ 45$ & $ 50$ & $ 92$ & & $   -$ & $1158$ & $1061$ & $1269$ & $1407$ & $1856$ \\\cline{7-7}\cline{14-14}
$6.57\times 10^{-2}$ & & $  -$ & $ 39$ & $ 35$ & $ 44$ & $ 41$ & $ 91$ & & $   -$ & $1116$ & $1074$ & $1255$ & $1170$ & $1833$ \\
$6.42\times 10^{-2}$ & & $  -$ & $ 38$ & $ 35$ & $ 44$ & $ 40$ & $ 91$ & & $   -$ & $1102$ & $1086$ & $1271$ & $1153$ & $1830$ \\\cline{6-6}\cline{13-13}
$6.26\times 10^{-2}$ & & $  -$ & $ 38$ & $ 35$ & $ 42$ & $ 40$ & $ 90$ & & $   -$ & $1115$ & $1099$ & $1232$ & $1167$ & $1807$ \\\cline{5-5}\cline{8-8}\cline{12-12}\cline{15-15}
$6.11\times 10^{-2}$ & & $  -$ & $ 38$ & $ 31$ & $ 41$ & $ 39$ & $ 99$ & & $   -$ & $1131$ & $ 986$ & $1218$ & $1150$ & $1997$ \\\cline{4-4}\cline{11-11}
$5.95\times 10^{-2}$ & & $  -$ & $ 34$ & $ 31$ & $ 40$ & $ 39$ & $ 96$ & & $   -$ & $1029$ & $ 997$ & $1203$ & $1161$ & $1934$ \\
$5.79\times 10^{-2}$ & & $  -$ & $ 33$ & $ 31$ & $ 40$ & $ 40$ & $ 94$ & & $   -$ & $1012$ & $1008$ & $1218$ & $1203$ & $1892$ \\\cline{7-7}\cline{14-14}
$5.64\times 10^{-2}$ & & $  -$ & $ 33$ & $ 31$ & $ 40$ & $ 37$ & $ 92$ & & $   -$ & $1026$ & $1018$ & $1233$ & $1128$ & $1851$ \\
$5.48\times 10^{-2}$ & & $  -$ & $ 32$ & $ 31$ & $ 40$ & $ 37$ & $ 89$ & & $   -$ & $1009$ & $1028$ & $1249$ & $1139$ & $1789$ \\\cline{3-3}\cline{10-10}
$5.33\times 10^{-2}$ & & $ 13$ & $ 32$ & $ 31$ & $ 39$ & $ 37$ & $ 87$ & & $ 737$ & $1024$ & $1038$ & $1234$ & $1150$ & $1749$ \\
$5.17\times 10^{-2}$ & & $ 13$ & $ 32$ & $ 31$ & $ 39$ & $ 37$ & $ 86$ & & $ 759$ & $1038$ & $1047$ & $1249$ & $1161$ & $1730$ \\
$5.02\times 10^{-2}$ & & $ 14$ & $ 33$ & $ 32$ & $ 39$ & $ 37$ & $ 85$ & & $ 843$ & $1085$ & $1090$ & $1267$ & $1171$ & $1711$ \\
$4.86\times 10^{-2}$ & & $ 15$ & $ 33$ & $ 32$ & $ 39$ & $ 37$ & $ 83$ & & $ 929$ & $1100$ & $1099$ & $1283$ & $1182$ & $1672$ \\
$4.71\times 10^{-2}$ & & $ 16$ & $ 33$ & $ 32$ & $ 39$ & $ 37$ & $ 82$ & & $1019$ & $1116$ & $1107$ & $1300$ & $1192$ & $1655$ \\\cline{8-8}\cline{15-15}
  \hline\hline
\end{tabular}
\caption{Proton and neutron number in the pasta phase region for
several nuclear interactions. The horizontal lines represent the shape
transitions, droplet-rod, rod-slab and slab-core. For NL3 only the droplet-core transition
occurs, and for IU-FSU the droplet-rod transition occurs at the lowest density shown.}\label{TabZNPat}
\end{table*}

\begin{table*}
  \centering
\begin{tabular}{c|cccccccccccccc}
  \hline\hline
 $\rho$ & &\multicolumn{6}{c}{$R_w\ [\mbox{fm}]$} & &\multicolumn{6}{c}{$E/A\ [MeV]$} \\
\cline{3-8}\cline{10-15}
   $[\mbox{fm}^{-3}]$& & NL3 & DD-ME2 & DD-ME$\delta$ & NL3$_{\omega\rho}$ & FSU & IU-FSU & &
   NL3 & DD-ME2 & DD-ME$\delta$ & NL3$_{\omega\rho}$ & FSU & IU-FSU \\
  \hline
$8.94\times 10^{-2}$ & &        &        &        &    $-$ &        &    $-$ & &          &          &          &$10.043$  &          & $10.945$ \\\cline{8-8}\cline{15-15}
$8.88\times 10^{-2}$ & &        &        &        &    $-$ &        & $23.6$ & &          &          &          &$10.000$  &          & $10.922$ \\
$8.35\times 10^{-2}$ & &        &        &        &    $-$ &        & $16.9$ & &          &          &          & $9.614$  &          & $10.693$ \\\cline{6-6}\cline{13-13}
$8.29\times 10^{-2}$ & &        &        &        & $11.7$ &        & $16.5$ & &          &          &          & $9.570$  &          & $10.667$ \\
$7.82\times 10^{-2}$ & & $ 9.2$ & $12.0$ & $10.5$ & $11.5$ & $16.1$ & $15.4$ & & $ 7.512$ & $ 9.351$ & $ 9.167$ & $ 9.217$ & $ 9.241$ & $10.448$ \\
$7.66\times 10^{-2}$ & & $ 8.0$ & $13.0$ & $ 7.0$ & $11.7$ & $ 2.0$ & $15.4$ & & $ 7.324$ & $ 9.230$ & $ 9.060$ & $ 9.100$ & $ 9.130$ & $10.373$ \\\cline{5-5}\cline{12-12}
$7.51\times 10^{-2}$ & & $ 8.8$ & $11.0$ & $10.7$ & $12.0$ & $10.2$ & $15.4$ & & $ 7.141$ & $ 9.110$ & $ 8.951$ & $ 8.983$ & $ 9.021$ & $10.296$ \\\cline{6-6}\cline{7-7}\cline{13-13}\cline{14-14}
$7.35\times 10^{-2}$ & & $ 9.3$ & $10.0$ & $10.8$ & $14.1$ & $11.8$ & $15.3$ & & $ 6.961$ & $ 8.989$ & $ 8.842$ & $ 8.865$ & $ 8.913$ & $10.217$ \\\cline{4-4}\cline{11-11}
$7.20\times 10^{-2}$ & & $ 7.9$ & $11.7$ & $11.0$ & $14.0$ & $11.5$ & $15.5$ & & $ 6.786$ & $ 8.868$ & $ 8.734$ & $ 8.746$ & $ 8.802$ & $10.137$ \\\cline{5-5}\cline{12-12}
$7.04\times 10^{-2}$ & & $ 8.9$ & $11.5$ & $13.1$ & $14.1$ & $11.5$ & $15.6$ & & $ 6.614$ & $ 8.745$ & $ 8.626$ & $ 8.627$ & $ 8.691$ & $10.055$ \\
$6.88\times 10^{-2}$ & & $ 8.4$ & $11.7$ & $13.1$ & $14.1$ & $11.6$ & $15.8$ & & $ 6.445$ & $ 8.622$ & $ 8.517$ & $ 8.507$ & $ 8.581$ & $ 9.972$ \\\cline{4-4}\cline{11-11}
$6.73\times 10^{-2}$ & & $ 9.0$ & $13.8$ & $13.1$ & $14.4$ & $12.0$ & $16.1$ & & $ 6.281$ & $ 8.498$ & $ 8.409$ & $ 8.386$ & $ 8.470$ & $ 9.886$ \\\cline{7-7}\cline{14-14}
$6.57\times 10^{-2}$ & & $ 9.3$ & $13.7$ & $13.4$ & $14.5$ & $14.0$ & $16.3$ & & $ 6.121$ & $ 8.373$ & $ 8.300$ & $ 8.264$ & $ 8.359$ & $ 9.798$ \\
$6.42\times 10^{-2}$ & & $ 8.7$ & $13.7$ & $13.6$ & $14.7$ & $14.0$ & $16.6$ & & $ 5.964$ & $ 8.248$ & $ 8.192$ & $ 8.142$ & $ 8.248$ & $ 9.708$ \\\cline{6-6}\cline{13-13}
$6.26\times 10^{-2}$ & & $ 9.4$ & $14.0$ & $13.9$ & $16.9$ & $14.3$ & $16.8$ & & $ 5.812$ & $ 8.123$ & $ 8.083$ & $ 8.019$ & $ 8.136$ & $ 9.616$ \\\cline{5-5}\cline{8-8}\cline{12-12}\cline{15-15}
$6.11\times 10^{-2}$ & & $ 9.3$ & $14.3$ & $15.8$ & $17.0$ & $14.4$ & $19.1$ & & $ 5.664$ & $ 7.998$ & $ 7.974$ & $ 7.896$ & $ 8.024$ & $ 9.522$ \\\cline{4-4}\cline{11-11}
$5.95\times 10^{-2}$ & & $ 9.6$ & $16.2$ & $16.0$ & $17.1$ & $14.6$ & $19.0$ & & $ 5.519$ & $ 7.871$ & $ 7.866$ & $ 7.771$ & $ 7.911$ & $ 9.423$ \\
$5.79\times 10^{-2}$ & & $ 9.9$ & $16.3$ & $16.2$ & $17.3$ & $15.1$ & $19.1$ & & $ 5.379$ & $ 7.745$ & $ 7.757$ & $ 7.646$ & $ 7.798$ & $ 9.322$ \\\cline{7-7}\cline{14-14}
$5.64\times 10^{-2}$ & & $19.3$ & $16.5$ & $16.4$ & $17.5$ & $17.0$ & $19.1$ & & $ 5.241$ & $ 7.617$ & $ 7.647$ & $ 7.520$ & $ 7.685$ & $ 9.219$ \\
$5.48\times 10^{-2}$ & & $20.6$ & $16.5$ & $16.6$ & $17.8$ & $17.2$ & $19.1$ & & $ 5.108$ & $ 7.490$ & $ 7.538$ & $ 7.394$ & $ 7.570$ & $ 9.112$ \\\cline{3-3}\cline{10-10}
$5.33\times 10^{-2}$ & & $15.0$ & $16.8$ & $16.9$ & $17.9$ & $17.5$ & $19.1$ & & $ 4.977$ & $ 7.362$ & $ 7.429$ & $ 7.266$ & $ 7.455$ & $ 9.003$ \\
$5.17\times 10^{-2}$ & & $15.3$ & $17.0$ & $17.1$ & $18.1$ & $17.7$ & $19.3$ & & $ 4.850$ & $ 7.234$ & $ 7.319$ & $ 7.138$ & $ 7.339$ & $ 8.890$ \\
$5.02\times 10^{-2}$ & & $16.0$ & $17.5$ & $17.5$ & $18.4$ & $17.9$ & $19.5$ & & $ 4.726$ & $ 7.105$ & $ 7.208$ & $ 7.010$ & $ 7.223$ & $ 8.774$ \\
$4.86\times 10^{-2}$ & & $16.7$ & $17.7$ & $17.7$ & $18.7$ & $18.2$ & $19.6$ & & $ 4.606$ & $ 6.976$ & $ 7.097$ & $ 6.880$ & $ 7.105$ & $ 8.654$ \\
$4.71\times 10^{-2}$ & & $17.4$ & $18.0$ & $17.9$ & $18.9$ & $18.4$ & $19.8$ & & $ 4.491$ & $ 6.847$ & $ 6.986$ & $ 6.750$ & $ 6.987$ & $ 8.531$ \\\cline{8-8}\cline{15-15}
  \hline\hline
\end{tabular}
\caption{Cell Wigner-Seitz radius and energy per baryon in the
pasta phase region.The horizontal lines represent the shape
transitions, droplet-rod, rod-slab and slab-core. For NL3 only the droplet-core transition
occurs, and for IU-FSU the droplet-rod transition occurs at the lowest density shown.}\label{TabRWEPat}
\end{table*}

\begin{table*}
  \centering
\begin{tabular}{c|cccccccccccccccc}
  \hline\hline
 $\rho$ & &\multicolumn{6}{c}{$\mu_N\ [MeV]$} & &\multicolumn{6}{c}{$x$} \\
\cline{3-8}\cline{10-15}
   $[\mbox{fm}^{-3}]$& & NL3 & DD-ME2 & DD-ME$\delta$ & NL3$_{\omega\rho}$ & FSU & IU-FSU & &
   NL3 & DD-ME2 & DD-ME$\delta$ & NL3$_{\omega\rho}$ & FSU & IU-FSU \\
  \hline
$8.94\times 10^{-2}$ & &          &          &          & $15.930$ &          & $13.711$ & &          &          &          & $ 0.040$ &          & $ 0.045$ \\\cline{8-8}\cline{15-15}
$8.88\times 10^{-2}$ & &          &          &          & $15.837$ &          & $13.941$ & &          &          &          & $ 0.040$ &          & $ 0.045$ \\
$8.35\times 10^{-2}$ & &          &          &          & $15.006$ &          & $13.752$ & &          &          &          & $ 0.038$ &          & $ 0.046$ \\\cline{6-6}\cline{13-13}
$8.29\times 10^{-2}$ & &          &          &          & $15.049$ &          & $13.730$ & &          &          &          & $ 0.038$ &          & $ 0.046$ \\
$7.82\times 10^{-2}$ & & $16.359$ & $14.759$ & $13.924$ & $14.419$ & $14.150$ & $13.529$ & & $ 0.029$ & $ 0.037$ & $ 0.035$ & $ 0.037$ & $ 0.037$ & $ 0.047$ \\
$7.66\times 10^{-2}$ & & $15.800$ & $14.518$ & $13.672$ & $14.207$ & $13.876$ & $13.460$ & & $ 0.028$ & $ 0.037$ & $ 0.034$ & $ 0.037$ & $ 0.037$ & $ 0.047$ \\\cline{5-5}\cline{12-12}
$7.51\times 10^{-2}$ & & $15.250$ & $14.278$ & $13.534$ & $13.995$ & $13.613$ & $13.387$ & & $ 0.027$ & $ 0.036$ & $ 0.034$ & $ 0.036$ & $ 0.036$ & $ 0.047$ \\\cline{6-6}\cline{7-7}\cline{13-13}\cline{14-14}
$7.35\times 10^{-2}$ & & $14.712$ & $14.037$ & $13.310$ & $13.822$ & $13.473$ & $13.312$ & & $ 0.027$ & $ 0.036$ & $ 0.033$ & $ 0.036$ & $ 0.036$ & $ 0.047$ \\\cline{4-4}\cline{11-11}
$7.20\times 10^{-2}$ & & $14.180$ & $13.867$ & $13.084$ & $13.613$ & $13.255$ & $13.235$ & & $ 0.026$ & $ 0.036$ & $ 0.033$ & $ 0.035$ & $ 0.036$ & $ 0.047$ \\\cline{5-5}\cline{12-12}
$7.04\times 10^{-2}$ & & $13.661$ & $13.641$ & $12.891$ & $13.402$ & $13.034$ & $13.155$ & & $ 0.025$ & $ 0.035$ & $ 0.033$ & $ 0.035$ & $ 0.035$ & $ 0.047$ \\
$6.88\times 10^{-2}$ & & $13.154$ & $13.412$ & $12.673$ & $13.189$ & $12.814$ & $13.073$ & & $ 0.024$ & $ 0.035$ & $ 0.032$ & $ 0.035$ & $ 0.035$ & $ 0.047$ \\\cline{4-4}\cline{11-11}
$6.73\times 10^{-2}$ & & $12.654$ & $13.216$ & $12.454$ & $12.976$ & $12.595$ & $12.989$ & & $ 0.023$ & $ 0.034$ & $ 0.032$ & $ 0.034$ & $ 0.034$ & $ 0.047$ \\\cline{7-7}\cline{14-14}
$6.57\times 10^{-2}$ & & $12.166$ & $12.987$ & $12.236$ & $12.759$ & $12.417$ & $12.902$ & & $ 0.023$ & $ 0.034$ & $ 0.032$ & $ 0.034$ & $ 0.034$ & $ 0.047$ \\
$6.42\times 10^{-2}$ & & $11.687$ & $12.756$ & $12.019$ & $12.543$ & $12.204$ & $12.813$ & & $ 0.022$ & $ 0.033$ & $ 0.031$ & $ 0.033$ & $ 0.034$ & $ 0.047$ \\\cline{6-6}\cline{13-13}
$6.26\times 10^{-2}$ & & $11.219$ & $12.521$ & $11.803$ & $12.338$ & $11.996$ & $12.718$ & & $ 0.021$ & $ 0.033$ & $ 0.031$ & $ 0.033$ & $ 0.033$ & $ 0.047$ \\\cline{5-5}\cline{8-8}\cline{12-12}\cline{15-15}
$6.11\times 10^{-2}$ & & $10.763$ & $12.287$ & $11.597$ & $12.119$ & $11.785$ & $12.679$ & & $ 0.020$ & $ 0.033$ & $ 0.030$ & $ 0.033$ & $ 0.033$ & $ 0.047$ \\\cline{4-4}\cline{11-11}
$5.95\times 10^{-2}$ & & $10.318$ & $12.064$ & $11.384$ & $11.896$ & $11.575$ & $12.584$ & & $ 0.020$ & $ 0.032$ & $ 0.030$ & $ 0.032$ & $ 0.033$ & $ 0.047$ \\
$5.79\times 10^{-2}$ & & $ 9.884$ & $11.825$ & $11.172$ & $11.673$ & $11.368$ & $12.485$ & & $ 0.019$ & $ 0.032$ & $ 0.030$ & $ 0.032$ & $ 0.032$ & $ 0.047$ \\\cline{7-7}\cline{14-14}
$5.64\times 10^{-2}$ & & $ 9.479$ & $11.585$ & $10.960$ & $11.447$ & $11.172$ & $12.382$ & & $ 0.018$ & $ 0.031$ & $ 0.030$ & $ 0.031$ & $ 0.032$ & $ 0.047$ \\
$5.48\times 10^{-2}$ & & $ 9.087$ & $11.344$ & $10.750$ & $11.219$ & $10.966$ & $12.274$ & & $ 0.018$ & $ 0.031$ & $ 0.029$ & $ 0.031$ & $ 0.031$ & $ 0.047$ \\\cline{3-3}\cline{10-10}
$5.33\times 10^{-2}$ & & $ 8.752$ & $11.102$ & $10.541$ & $10.990$ & $10.760$ & $12.162$ & & $ 0.017$ & $ 0.030$ & $ 0.029$ & $ 0.031$ & $ 0.031$ & $ 0.047$ \\
$5.17\times 10^{-2}$ & & $ 8.370$ & $10.857$ & $10.333$ & $10.757$ & $10.555$ & $12.045$ & & $ 0.017$ & $ 0.030$ & $ 0.029$ & $ 0.030$ & $ 0.031$ & $ 0.047$ \\
$5.02\times 10^{-2}$ & & $ 7.999$ & $10.612$ & $10.127$ & $10.524$ & $10.349$ & $11.922$ & & $ 0.016$ & $ 0.030$ & $ 0.029$ & $ 0.030$ & $ 0.031$ & $ 0.047$ \\
$4.86\times 10^{-2}$ & & $ 7.637$ & $10.365$ & $ 9.922$ & $10.288$ & $10.144$ & $11.793$ & & $ 0.016$ & $ 0.029$ & $ 0.028$ & $ 0.030$ & $ 0.030$ & $ 0.047$ \\
$4.71\times 10^{-2}$ & & $ 7.285$ & $10.118$ & $ 9.717$ & $10.050$ & $ 9.938$ & $11.659$ & & $ 0.015$ & $ 0.029$ & $ 0.028$ & $ 0.029$ & $ 0.030$ & $ 0.047$ \\\cline{8-8}\cline{15-15}
  \hline\hline
\end{tabular}
\caption{Neutron chemical potential and proton fraction in the
pasta phase region.The horizontal lines represent the shape
transitions, droplet-rod, rod-slab and slab-core. For NL3 only the droplet-core transition
occurs, and for IU-FSU the droplet-rod transition occurs at the lowest density shown.}\label{TabMuXPat}
\end{table*}

\begin{table*}
  \centering
  \begin{tabular}{c|ccccccccc}
    \hline\hline
    Transition & & SLy4(HF) & SLy4(TF) & NL3 & DD-ME2 & DD-ME$\delta$ & NL3$_{\omega\rho}$ & FSU & IU-FSU \\
    \hline
    droplet - rod      & & 0.042 & 0.066 &      - & 0.0611 & 0.0626 & 0.0642 & 0.0580 & 0.0471 \\
    rod - slab         & & 0.070 & 0.078 &      - & 0.0688 & 0.0720 & 0.0751 & 0.0673 & 0.0626 \\
    slab - homogeneous & & 0.080 & 0.085 & 0.0548$^*$ & 0.0735 & 0.0766 & 0.0835 & 0.0751 & 0.0894 \\
    \hline\hline
  \end{tabular}
  \caption{Densities at the shape transition. The densities are give in fm$^{-3}$.
  SLy4(HF) and SLy4(TF) refer to the results found in \cite{Gogelein}. For
  NL3, only the transition droplet-core is indicated.}\label{TabTra}
\end{table*}

\begin{figure*}
\begin{tabular}{cc}
  \includegraphics[width=0.5\linewidth]{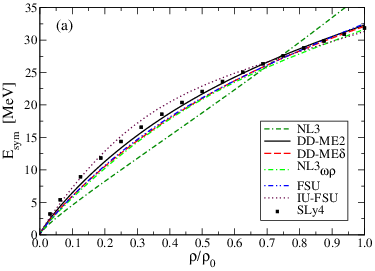}&
  \includegraphics[width=0.5\linewidth]{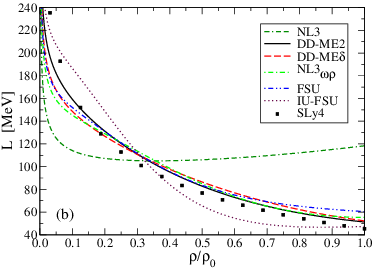}\\
\end{tabular}
  \caption{(Colors online) Symmetry energy $E_{sym}$ (a) and its slope $L$ (b)
 as a function of the density $\rho$.}
\label{EsymFig}
\end{figure*}

\begin{figure}
  \includegraphics[width=1.0\linewidth]{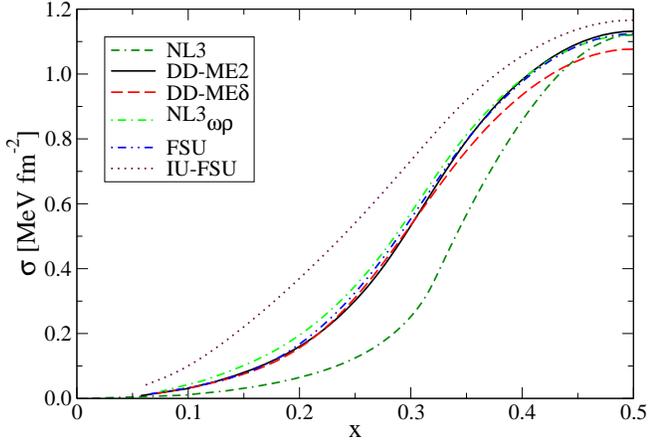}
  \caption{(Colors online)
Surface tension as a function of the proton fraction x for all the models
considered in this study except SLy4.
}
\label{sigma}
\end{figure}

\begin{figure}
\begin{tabular}{c}
  \includegraphics[width=1.0\linewidth]{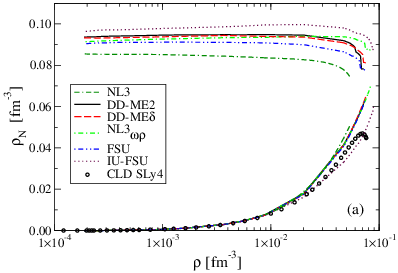}\\
  \includegraphics[width=1.0\linewidth]{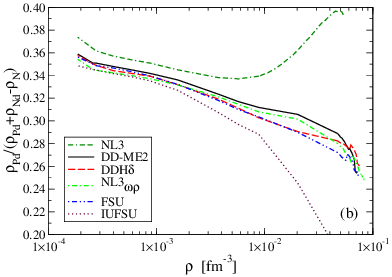}\\
  \includegraphics[width=1.0\linewidth]{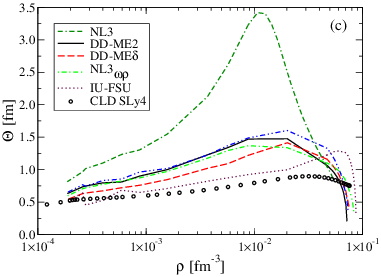}\\
\end{tabular}
  \caption{(Colors online) Neutron density at the cell  center and border (a),
    cluster proton fraction at the cluster center (b) and neutron skin thickness (c)
for all the models considered in this study.}
\label{dnddx}
\end{figure}

\begin{figure*}
\begin{tabular}{cc}
  \includegraphics[width=0.5\linewidth]{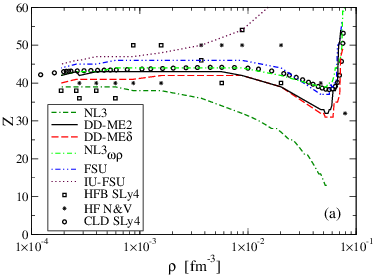}&
  \includegraphics[width=0.5\linewidth]{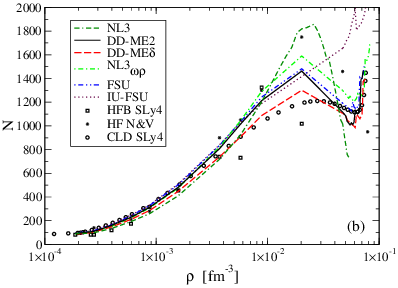}\\
  \includegraphics[width=0.5\linewidth]{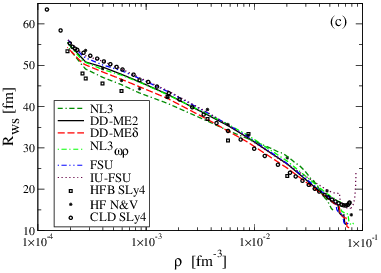}&
  \includegraphics[width=0.5\linewidth]{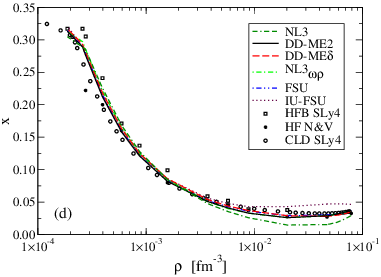}\\
  \includegraphics[width=0.5\linewidth]{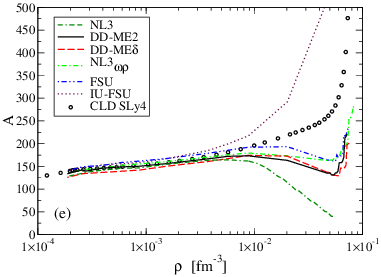}&
  \includegraphics[width=0.5\linewidth]{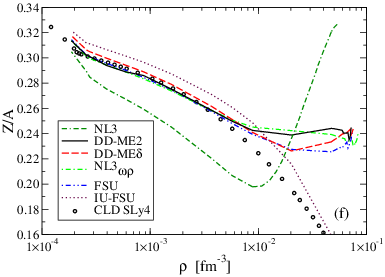}\\
\end{tabular}

  \caption{(Colors online) Proton number $Z$ (a), neutron number $N$ (b), Wigner-Seitz cell radius
    $R_{ws}$ (c), proton fraction $x$ (d), the baryon number in the nucleus $A$ (e) and the proton fraction in the nucleus
    $Z/A$ (f) as a function of the  density ($\rho$).
  HFB SLy4, HF $N\&V$ and CLD $SLy4$ refer respectively to the results found in~\cite{grill},~\cite{NV} and~\cite{Haensel}.}\label{ZNRwXFig}
\end{figure*}

\begin{figure}
  \includegraphics[width=1.0\linewidth]{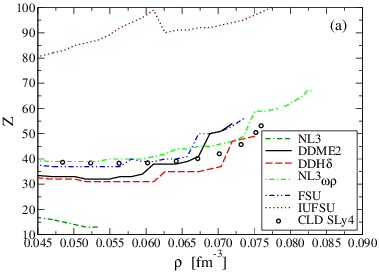}\\
  \includegraphics[width=1.0\linewidth]{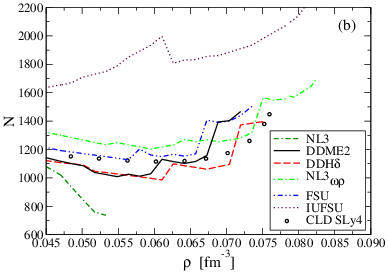}\\
  \includegraphics[width=1.0\linewidth]{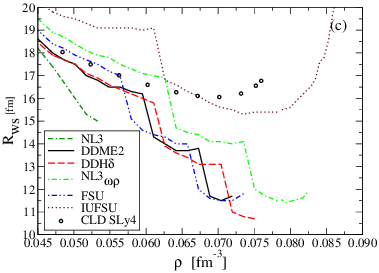}
  \caption{(Colors online) Proton number (a), neutron number (b) and Wigner-Seitz cell radius (c) in the
    pasta phase regions. CLD $SLy4$ refers to the results found in~\cite{Haensel}.}
\label{ZNRwFigPat}
\end{figure}

\end{document}